\documentclass[a4paper,11pt]{llncs}
\usepackage{amsmath}
\usepackage{pstricks}
\usepackage{graphicx}
\usepackage[latin1]{inputenc}
\usepackage{ifthen}
\usepackage{url}

\newcommand{\colorfigs}{\def\@colorfigs{true}}
\colorfigs{}

\title{Seed design framework for mapping SOLiD reads}
\author{Laurent No\'e \and Marta G\^irdea \and Gregory Kucherov\thanks{On leave in J.-V.Poncelet Lab, Moscow, Russia}}
\institute{INRIA Lille - Nord Europe, LIFL/CNRS, Universit\'e Lille 1, 59655 Villeneuve d'Ascq, France}

\begin{document}

\maketitle

\begin{abstract}
The advent of high-throughput sequencing technologies constituted a major advance in
genomic studies, offering new prospects in a wide range of
applications. We propose a rigorous and flexible algorithmic solution 
to mapping SOLiD color-space reads to a reference genome. The solution
relies on an advanced method of seed design that uses a faithful
probabilistic model of read matches and, on the other hand, a novel
seeding principle especially adapted to read mapping. Our method can
handle both lossy and lossless frameworks and is able to distinguish,
at the level of seed design, between SNPs and reading errors. 
We illustrate our approach by several seed designs and demonstrate
their efficiency. 
\end{abstract}

\section{Introduction}

High-throughput sequencing technologies can produce hundreds of millions of DNA sequence
reads in a single run, providing faster and less expensive solutions to a wide range
of genomic problems. Among them, the popular SOLiD system (Applied Biosystems)
features a 2-base encoding of reads, with an error-correcting
capability helping to reduce the error rate and to better
distinguish between sequencing errors and SNPs. 

In this paper, we propose a rigorous and flexible algorithmic approach
to mapping SOLiD color-space reads to a reference genome, capable to
take into account various external parameters as
well as intrinsic properties of reads resulting from the SOLiD
technology. The flexibility and power of our approach comes from an
advanced use of {\em spaced seeds}
\cite{PatternHunter02,noe2005yass}.

The main novelty of our method is an {\em advanced
  seed design} based on a {\em faithful probabilistic model of SOLiD read
  alignments} incorporating reading errors, SNPs and base indels,
 and, on the other hand, on a {\em new seeding principle}
especially adapted for read mapping.
The latter relies on
the use of a small number of seeds (in practice, typically two)
{\em designed simultaneously with a set of position on the read where they
  can hit}. We call this principle {\em position-restricted
  seeds}. Advantageously, it allows us to take into account, in a
subtle way, read properties such as a non-uniform
distribution of reading errors along the read, or a tendency of
reading errors to occur periodically at a distance of 5 positions,
which are observed artifacts of the SOLiD technology.

A number of algorithms and associated software programs for read
mapping have been recently published. Several of them such as
MAQ~\cite{li2008psd},
MOSAIK~\cite{mosaikweb},
MPSCAN~\cite{rivals2009mpscan}
PASS~\cite{campagna2009pas},
PerM~\cite{chen2009perm},
RazerS~\cite{weese2009razers},
SHRiMP~\cite{10.1371/journal.pcbi.1000386}
or
ZOOM \cite{lin2008zzo}
apply contiguous or spaced seeding techniques, requiring one or several hits per read.
Other programs approach the problem differently,
e.g., by using the Burrows-Wheeler transform (Bowtie~\cite{langmead2009ume},
BWA~\cite{li2009fast}, SOAP2 \cite{soap2}),
suffix arrays (segemehl~\cite{hoffmann2009fast}, BFAST~\cite{homer09bfast}),
variations of the Rabin-Karp algorithm (SOCS~\cite{ondov2008efficient})
or a non-deterministic automata matching algorithm on a
keyword tree of the search strings (PatMaN~\cite{prufer2008patman}).
Some tools, such as segemehl~\cite{hoffmann2009fast}
or Eland~\cite{bentley2008accurate},
are designed for 454 and Illumina reads 
and thus do not deal with the characteristics of the SOLiD encoding which is the subject of this paper.
Also, it should be noted that, in many cases, sensitivity is sacrificed
in favor of speed: most methods find similarities up to a small number
of mismatches, and few approaches 
account for nucleotide insertions and deletions.

Seed-based methods for read mapping use
different seeding strategies. SHRiMP
\cite{10.1371/journal.pcbi.1000386} 
uses spaced seeds that can hit at any position of the read and introduces a
lower bound on the number of hits within one read.
MAQ~\cite{li2008psd} uses six light-weight seeds allowed to hit in the
initial part of the read. ZOOM
\cite{lin2008zzo} 
proposes to use a small number (4-6) of
spaced seeds each applying at a fixed position,
to ensure a lossless search with respect to a given
number of mismatches. In the lossless framework, PerM
\cite{chen2009perm} proposes to use ``periodic seeds'' (see also
\cite{kucherov2005mlf}) to save on the index size. 

Despite the number of proposed solutions, none of them relies on a
systematic seed design method taking into account (other than very
empirically) statistical properties of reads. In this paper, we
present a seed design based on Hidden Markov models of read matches, using
a formal finite automata-based approach previously developed in
\cite{kucherov2006ufs}. To the best of our knowledge, this is the
first time that the seed design for read mapping is done based on a
rigorous probabilistic modeling.  

Our approach allows us to design seeds
in both lossy and lossless frameworks. In the lossless framework,
where the goal is to detect all read occurrences within a specified
number of mismatches, we have the flexibility of partitioning this number
into reading errors and SNPs.

As a result, we obtain a very efficient mapping algorithm combining a
small number of seeds and therefore a reasonable amount of index memory
with guaranteed sensitivity and small running time, due to a
restricted subset of positions where seeds should be applied.

\section{AB SOLiD reads: encoding and technological artifacts}\label{sec:absolid}
The SOLiD System~\cite{absolid1} enables massively parallel sequencing of
clonally amplified DNA fragments.
This sequencing technology is based on sequential ligation
of dye-labeled oligonucleotide probes, each probe
assaying two base positions at a time. The system uses four
fluorescent dyes to encode for the sixteen possible 2-base
combinations. Consequently, a DNA fragment is represented by
the initial base followed by a sequence of overlapping dimers,
each encoded with one of four colors using a degenerate coding
scheme that satisfies several rules.
Thus, although a single color in a read can represent any of four dimers,
the overlapping properties of the dimers and the nature
of the color code eliminate ambiguities and allow for
error-correcting properties.

As our work relies on modeling the error distribution along the reads,
we are particularly interested in several aspects of the sequencing
technology that influence this distribution. First,
since every color of the read encodes two adjacent bases
and therefore every base affects two adjacent colors,
it follows that any single base mutation results in
the change of two adjacent colors in the read.
On the other hand, since cycles of five di-nucleotide readings are performed
in order to retrieve the sequence
(as described in the documentation of Applied Biosystems~\cite{absolid1,absolid2}),
we expect reading error bias to appear with a periodicity of 5.

To confirm this intuition, we studied the variation of the reading
error probability along the read by analyzing statistical properties of about a million of SOLiD
reads of the {\em S. cerevisiae} genome. 
In this analysis, we used the qualities $Q_l$ associated 
to each position $l$ on the read, which relate to 
the error probability $p_e^l$ through
$Q_l = -10 \cdot \log_{10}{(p_e^l)}$~\cite{ewing1998bcae}.

\begin{figure}[!hbt]
\begin{minipage}[t]{0.48\textwidth} 
\begin{center}
\hspace{0mm}\includegraphics[width=1\textwidth]{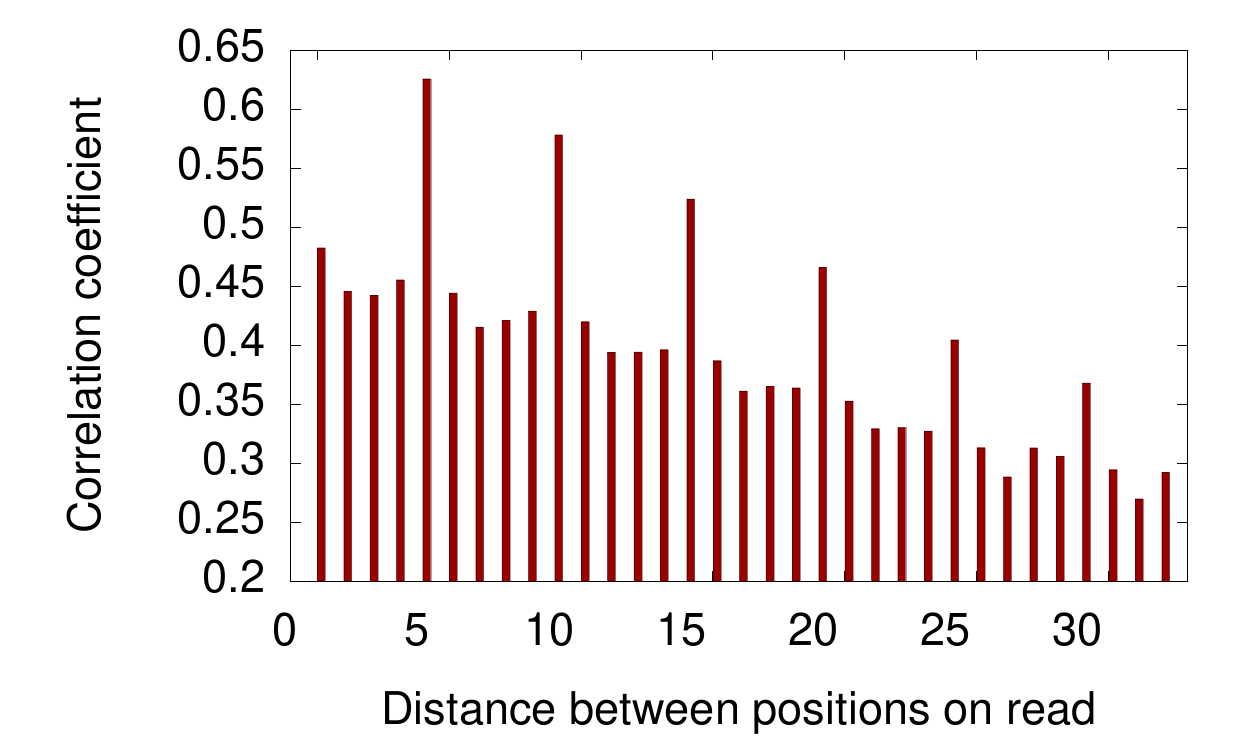}
\vspace{-8mm}
\end{center}
\caption{Position quality correlation coefficient depending
   on the distance between read positions.}
\label{fig:relative-err-pb}
\end{minipage}
\hspace{0.5em}
\begin{minipage}[t]{0.48\textwidth}
\begin{center}
\hspace{0mm}\includegraphics[width=1\textwidth]{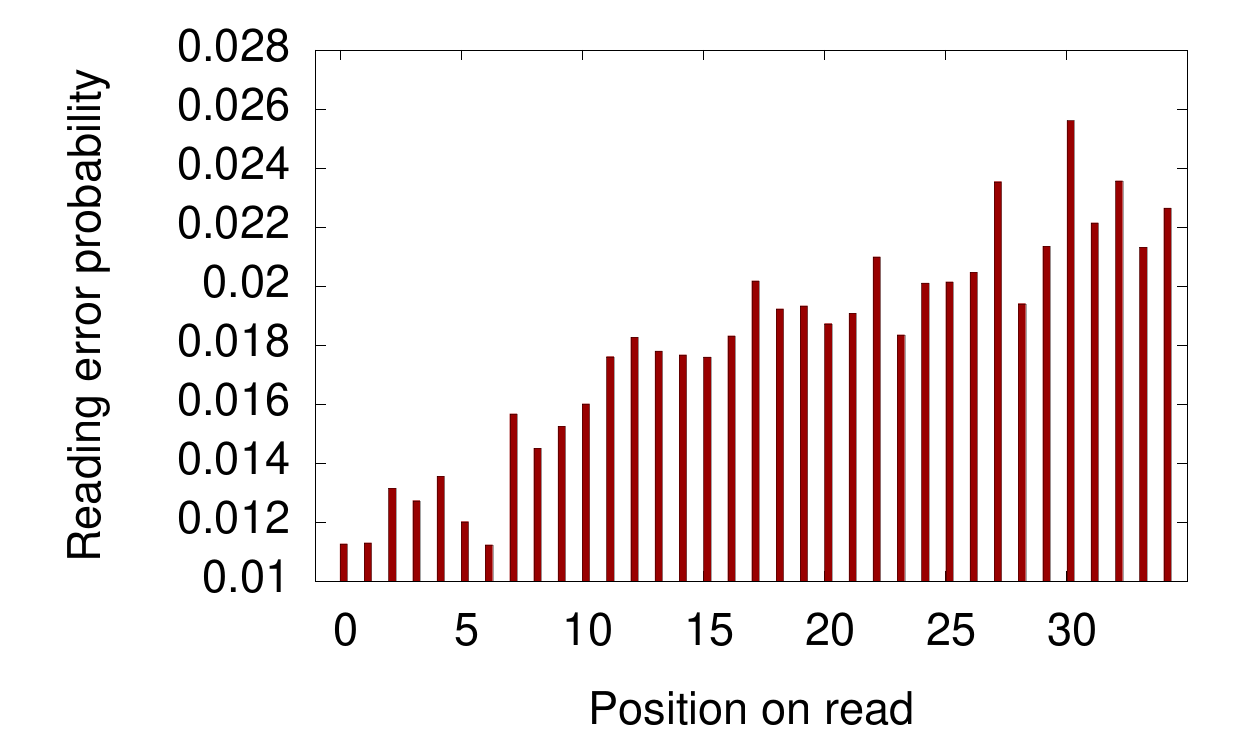}
\vspace{-8mm}
\end{center}
\caption{Average reading error probability at each read position.}
\label{fig:absolute-err-pb}
\end{minipage}
\end{figure}

We computed the quality correlation between read positions
depending on the distance between them. Formally, if
$m$ is the read length, then for
each $i \in \{1,..,m-1\}$, we computed the correlation through the
following standard formula
$c(i)  =\frac{E((Q_{j}-\widetilde{Q})(Q_{j+i}-\widetilde{Q}))}{(\sigma_{Q})^2}$,
where $E(\cdot)$ is the expectation, $\widetilde{Q}$ the average
quality along the read, and $\sigma_{Q}$ the standard deviation
of quality values. 
The result is given in Figure~\ref{fig:relative-err-pb}. 
It shows significantly higher correlations (up to 0.63)
between pairs of positions located at distances that are multiples of~5.

Additionally, we studied the behavior of reading error probability values along
the read. As shown in Figure~\ref{fig:absolute-err-pb},
the error probability tends to increase towards the end of the read,
making the last positions of the color sequence less reliable when
searching for similarities.

\section{Seed design for mapping SOLiD reads}\label{sec:seed-design}

\subsection{Seed design: background}\label{sec:seed-automaton}

Spaced seeds, first proposed in the context of DNA sequence alignment
by the PatternHunter algorithm \cite{PatternHunter02}, represent a
powerful tool for enhancing the efficiency of the sequence search.

Using a spaced seed instead of a contiguous stretch of identical nucleotides to
select a potential similarity region can improve the sensitivity of the
search for a given selectivity level \cite{PatternHunter02}. Furthermore, using 
a seed family, i.e. several seeds simultaneously instead of a single seed, further improves
the sensibility/selectivity trade-off \cite{PatternHunter04,SunBuhlerJCB05}. 
The price for using seed
families is the necessity to store in memory
several indexes, one for each seed. In practice, however, using in the
search a small number of seeds can significantly improve
the sensitivity/selectivity ratio. 

A crucial feature of spaced seeds is their capacity to be adapted to
different search situations. Spaced seeds can be {\em designed} to
capture statistical properties of sequences to be searched. For example,
\cite{brejova2003optimal,zhou2008universal} report on designing spaced
seeds adapted to the search of coding regions. One of the
contributions of this paper is a rigorous design of seeds adapted to
mapping genomic reads issued from the SOLiD technology.
Note that here we will work with regular spaced seeds rather than more advanced
subset seeds~\cite{kucherov2006ufs,zhou2008universal,YangZhangJCB08}, as there is very little or no information in discriminating among different classes of mismatches that can be used to our advantage.

One has to distinguish between the {\em lossy} and {\em lossless} cases of
seed-based search. In the lossy case we are allowed to miss a
fraction of target matches, and the usual goal of seed design is
to maximize the sensitivity over a class of seeds verifying a certain
selectivity level. In the lossless case we must detect all 
matches verifying a given dissimilarity
threshold (expressed in terms of a number of errors or a minimal score),
and the goal of seed design is to compute a minimal set of
seeds with the best selectivity that still ensures the lossless
search. In the context of read mapping for high-throughput sequencing
technologies, both lossy \cite{10.1371/journal.pcbi.1000386,li2008psd} and lossless
\cite{lin2008zzo,chen2009perm} frameworks have been used. 

Our approach to seed design relies on a methodology proposed in our
previous work \cite{kucherov2006ufs}, based on the finite automata
theory. A central idea is to model the set of target alignments by a
{\em finite-state probability transducer}, which subsumes the Hidden
Markov Model commonly used in biosequence analysis. On the other hand,
a seed, or a seed family, is modeled by a {\em seed automaton} for
which we proposed an efficient compact construction
\cite{kucherov2007ssa}. Once these two automata have been specified,
computing the seed sensitivity can be done efficiently with a dynamic
programming algorithm as described in \cite{kucherov2006ufs}. The seed
design is then done by applying
our {\sc Iedera} software~\cite{kucherov2006ufs,kucherov2007ssa,iederaweb}
that uses the above algorithm to explore the
space of possible seeds and select most sensitive seeds using a sampling
procedure for seeds and respective hit positions
and by performing a local optimization on the best candidates.

Here we apply this methodology to seed design for mapping SOLiD
reads, both in the lossy and lossless frameworks. Besides, we
introduce an important novelty in the definition of seeds, especially
advantageous for mapping short reads:
{\em position-restricted seeds}, which are seeds
designed together with the set of positions on the read where they can be applied.
This can be seen as
an intermediate paradigm between applying each
seed at every position and the framework of
\cite{lin2008zzo} where each seed
applies to a designated position of the read. 
Position-restricted seeds offer an additional power of capturing
certain read properties (such as, e.g., an increasing error level towards the
end of the read) in a flexible way, without sacrificing the
selectivity and thus the speed of the seeding procedure.

\subsection{Modeling seeds and SOLiD reads by finite automata}\label{sec:solid-fw}

We now present our model of color sequence alignments, built on the observations
of Section~\ref{sec:absolid}. 
Note that we consider the reference genome translated into the color alphabet,
i.e. both the reads and the genome are represented in color space. 

\subsubsection{Position-restricted seeds.}\label{sec:seed-pos}

As shown in Section~\ref{sec:absolid}, the reading error probability
increases towards the end of the read,
implying that a search for similarity within the last
positions of the read could lead to erroneous results or no results at all.
Hence, we can improve the seed selectivity by 
favoring hits at initial positions of the read
where matches are more likely to be significant.
We then define each seed $\pi$ {\em jointly} with a set of
positions $P$ to which it is applied on the read.

We use the framework of \cite{kucherov2006ufs} where a
seed $\pi$ is represented by a deterministic finite automaton
$\mathcal{Q}$ over the alignment alphabet $\mathcal{A}$
which is here the binary match/mismatch alphabet. 
Note that the size of $\mathcal{Q}$ is a crucial parameter
in the algorithm of \cite{kucherov2006ufs} to compute the sensitivity
of the seed. 
An efficient construction of such an automaton has been
studied in \cite{kucherov2007ssa}: it has the optimal size of
$(w+1)2^{s-w}$ states, where $s$ and $w$ are respectively the {\em
  span} (length) and {\em weight} (number of {\em match} symbols) of the
seed.

Let $m$ be the read size. To take into account the set of
allowed positions, we compute the product of $\mathcal{Q}$ 
with an automaton $\lambda_P$ consisting of a linear chain of
$m+1$ states $q_0,q_1,\ldots,q_m$, where $q_0$ is the initial state,
and for every $q_i$, both outgoing transitions lead to
$q_{i+1}$. Final states of the automaton reflect the set of possible
positions $P$ where the seed is allowed to hit: a state $q_i$ is final iff
$i-s\in P$.

A trivial upper bound on the size of the product automaton for 
a spaced seed of span $s$ and weight $w$ is $(w+1)
\cdot 2^{s-w} \cdot m$.
This bound can be improved using the notion of matching prefix,
as explained in \cite{kucherov2007ssa}. Thus, an
economical implementation of the product of $\mathcal{Q}$ by $\lambda$ 
taking into account the set of matching positions $P$ always produces at most
$(w+1) \cdot 2^{s-w} \cdot |P| + m$ states.

Furthermore, consider an interval graph of the possible
placements of the seed on the read,
where each placement spans over an interval of $s$ positions. The chromatic number
$c$ of this graph can be easily computed, providing the maximal
number of overlapping seeds. We observe that if this number is small 
(compared to $(s-w + log(w))$), then the size of the product automaton
is bounded by $O((m+1) \cdot 2^c)$.

\subsubsection{Model for SNPs and reading errors}\label{sec:snp-re}

As explained in Section~\ref{sec:absolid}, there are two independent
sources of errors in reads with respect to the reference genome:
reading errors and SNPs/indels, i.e., {\em bona fide} differences between 
the reference genome and sequenced data. We represent each of these
sources by a separate
Hidden Markov Model (viewed as a probabilistic transducer, see \cite{kucherov2006ufs}),
combined in a model which allows all error types to be cumulated
in the resulting sequences.

\begin{center}
\begin{figure}[!hbt]
\vspace{-2em}
\begin{minipage}[c]{0.58\textwidth} 
\begin{center}
\ifthenelse{\equal{\@colorfigs}{true}}{
\hspace{-1.7em}\includegraphics[width=\textwidth]{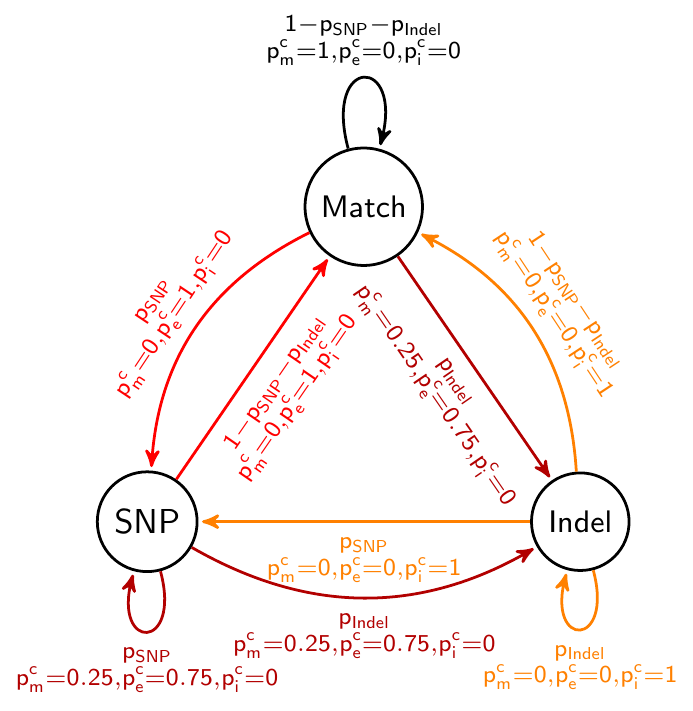}
}{
\hspace{-1.7em}\includegraphics[width=\textwidth]{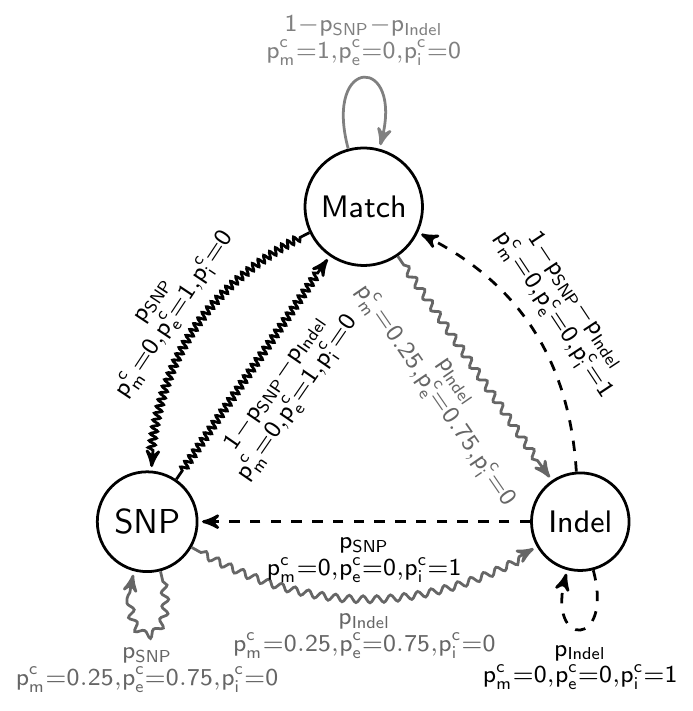}
}
\end{center}
\ifthenelse{\equal{\@colorfigs}{true}}{
\caption{Model of SNPs and Indels ($M_{SNP/I}$). Colors of transitions correspond
to emitted errors: black for color matches, red for mismatches, yellow for indels,
and dark red for a mixture of matches (0.25) and mismatches (0.75)}
}{
\caption{Model of SNPs and Indels ($M_{SNP/I}$). Colors and shapes of transitions correspond
to emitted errors: plain grey for color matches, dashed for indels, wavy black for mismatches,
wavy grey for a mixture of matches (0.25) and mismatches (0.75)}
}
\label{fig:err-distrib-automaton1}
\end{minipage}
\hspace{3mm}
\begin{minipage}[c]{0.4\textwidth}
\begin{center}
\ifthenelse{\equal{\@colorfigs}{true}}{
\hspace{-2em}\includegraphics[width=\textwidth]{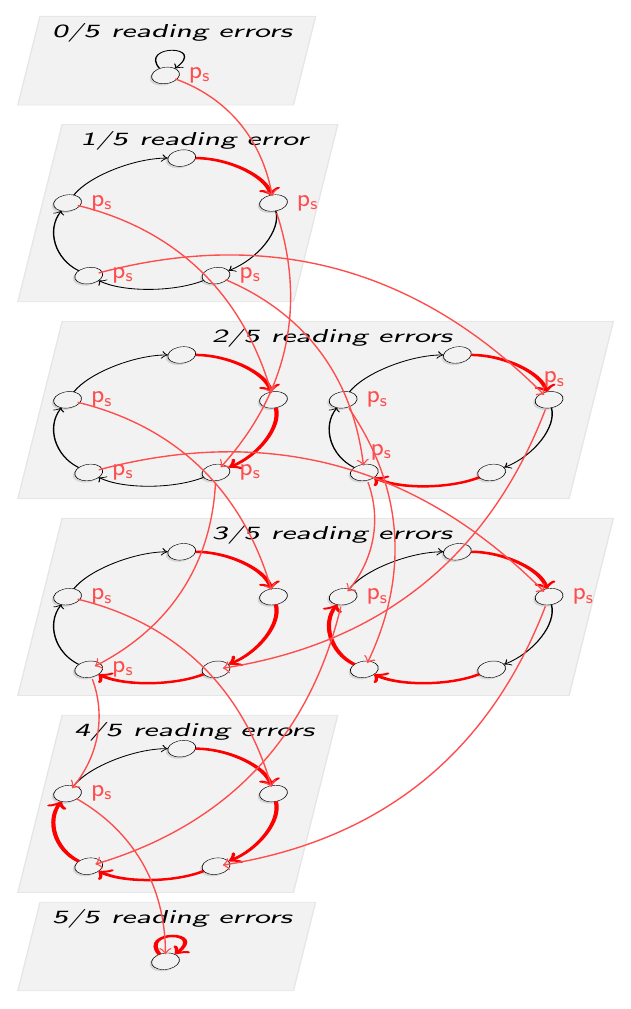}
}{
\hspace{-2em}\includegraphics[width=\textwidth]{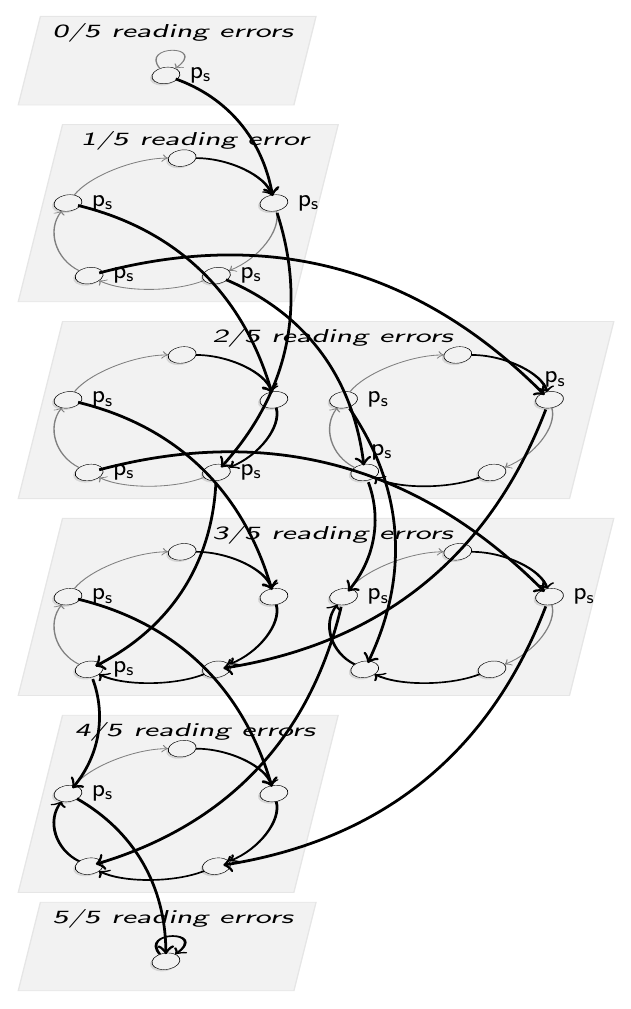}
}
\vspace{-3mm}
\end{center}
\caption{Reading error automaton}
\label{fig:err-distrib-automaton2}
\end{minipage}
\end{figure}
\end{center}

The {\bf SNP/Indel model}, denoted $M_{SNP/I}$, (Figure~\ref{fig:err-distrib-automaton1})
has three states: {\em Match}, {\em SNP} and {\em Indel},
referring to matches, mismatches, and indels {\em at the nucleotide level},
and is parametrized by SNP and Indel occurrence probabilities, denoted $p_{SNP}$ and $p_{Indel}$.
Each transition of $M_{SNP/I}$ generates a {\em color match, mismatch or indel}, with probabilities
$p^c_{m}$, $p^c_{e}$, and $p^c_{i}$ respectively, defined as follows.
An insertion or deletion of $n$ nucleotides appears at the color level as an
insertion/deletion of $n$ colors preceded in 3/4 cases by a color mismatch~\cite{absolid1}.
Hence, the $p^c_{e} = 0.75$ when entering the {\em Indel} state, and $p^c_{i} = 1$ for
any transition having the {\em Indel} state as source.
A nucleotide mutation is reflected in the color encoding by a change of two
adjacent colors (and, more generally, $n$ consecutive mutations
affect $n+1$ consecutive colors~\cite{absolid1}). Thus, $p^c_{e} = 1$
when entering or leaving the {\em SNP} state, and a color match/mismatch mixture 
when staying in the mismatch state, since color matches may occur inside stretches of consecutive SNPs.
Finally, $p^c_{m} = 1$ when looping on the {\em M} state.

The {\bf reading errors} are handled by a more complex model, 
denoted $M_{RE}$ (Figure~\ref{fig:err-distrib-automaton2}).
Basically, it is composed of several submodels, one for each 
possible arrangement of reading errors on a cycle of 5 positions.
\ifthenelse{\equal{\@colorfigs}{true}}{
Within these submodels, the transitions shown in red correspond to
}{
Within these submodels, the transitions shown in thick black correspond to
}
periodic reading errors, and
generate reading errors with a fixed, usually high probability
$p_{err}$. This simulates the periodicity property shown in
Figure~\ref{fig:relative-err-pb}. 
Switching from one cyclic submodel to another with a higher reading error rate
\ifthenelse{\equal{\@colorfigs}{true}}{
(by adding another red transition, with high error probability)
}{
(by adding another transition with high error probability)
}
can occur at any moment with a fixed probability~$p_s$.

\ifthenelse{\equal{\@colorfigs}{true}}{
The transitions shown in black in the model from Figure~\ref{fig:err-distrib-automaton2}
}{
The transitions shown in grey in the model from Figure~\ref{fig:err-distrib-automaton2}
}
have an error emission probability of 0. However, in the complete reading error
model, we wish to simulate the error probability that increases towards the end
(in conformity with Figure~\ref{fig:absolute-err-pb}).
We do this by ensuring that reading errors are generated on these transitions 
with a probability $p'_{err}(pos)$ (lower than $p_{err}$) given by an
increasing function of the current position $pos$ on the read.
Technically, this is achieved by multiplying
the automaton in Figure~\ref{fig:err-distrib-automaton2} by a linear
automaton with $m+1$ states, where $m$ is the read length and the $i$-th transition
generates a reading error (color mismatch) with the probability $p'_{err}(i)$.
The reading error emission probability in the product model is computed
as the maximum of the two reading error probabilities encountered
in the multiplied models.

The {\bf final model}, which combines both error sources,
is the product of $M_{SNP/I}$ and  $M_{RE}$. 
While the states and transitions of the product model are defined in the classic manner,
the emissions are defined through specific rules based on symbol priorities. 
If corresponding transitions of $M_{SNP/I}$ and $M_{RE}$ generate
symbols $\alpha$ and $\beta$ with probabilities $p_1$ and $p_2$
respectively, then the product automaton generates the dominant symbol
between $\alpha$ and $\beta$ with probability $p_1p_2$. Different
probabilities obtained in this way for the same symbol are added up. 

The dominance relation is defined as follows:
{\em indels} are dominant over both {\em mismatches} and {\em matches},
and {\em mismatches} dominate {\em matches}.
For example,
$(indel,mismatch)$ results in an $indel$,
$(mismatch, mismatch)$ and $(match, mismatch)$ represent $mismatch$,
$(match,match)$ is a $match$.
This approach ensures that errors generated by each of the two models
are superposed. 

\subsection{Computing the sensitivity or testing the lossless property}\label{sec:eval-methods}

Given an automaton $\mathcal{Q}$ specifying a family of seeds possibly
restricted to a set of positions, we have to compute its sensitivity
(in the lossy framework) or to test whether it is lossless (in the
lossless framework). 

The sensitivity of a seed family is defined~\cite{PatternHunter02,KeichLiMaTrompDAM04} as the
probability for at least one of the seeds to hit a read alignment with respect to a given
probabilistic model of the alignment. 
As outlined in Section~\ref{sec:seed-automaton}, this is done
using the dynamic programming technique of \cite{kucherov2006ufs}. We therefore omit
further details. 

In the lossless framework, we have to test if the seed specified by
$\mathcal{Q}$ is lossless, i.e. hits 
{\em all} the target alignments. 
The set of target alignments is defined through a threshold number of
allowed mismatches. 

A straightforward way to test the lossless property of $\mathcal{Q}$
would be to construct a deterministic automaton recognizing the set of all
target alignments and then to test if the language of this automaton is
included in the language of $\mathcal{Q}$. This, however, is
unfeasible in practice. The automaton of all
target alignments is much too costly to construct: for example, in the case
of threshold of $k$ mismatches, there are $\sum_{a=0}^{k}
\binom{m}{a}$ different alignments of length $m$, and the Aho-Corasick automaton of
these strings would have $\sum_{a=0}^{k+1} \binom{m}{a}$
states. Moreover, testing the inclusion would lead to computing the
product of this automaton with $\mathcal{Q}$ which would multiply
the number of states by that of $\mathcal{Q}$.

Alternatively, we propose an efficient dynamic programming algorithm directly applied to $\mathcal{Q}$ 
that can verify the {inclusion}. This algorithm computes, for each
state $q$ of $\mathcal{Q}$, and for each iteration $i \in [1..m]$, the minimal number
of mismatches needed to
reach $q$ at step $i$. Let $k$ be the threshold for the number of
mismatches. Then, the lossless condition holds iff at step $m$, all
non-final states have a number of mismatches greater than $k$. 
Indeed, if there is a non-final
state that has a number of errors
at most $k$ after $m$ steps,
then there is at least one string of length $m$ with
at most $k$ mismatches that is not detected by the
automaton, which contradicts the lossless condition. 
This algorithm is of time complexity $\mathcal{O}(|\mathcal{Q}| \cdot
|\mathcal{A}| \cdot m)$, and space complexity $\mathcal{O}(|\mathcal{Q}| \cdot
|\mathcal{A}|)$, 
where $\mathcal{A}$ is the alphabet of the alignment sequences, in our case $\{0, 1\}$ 

To illustrate the efficiency of this algorithm, consider the case
of a single spaced seed of span $s$ and weight $w$, yielding an
automaton with at most $(w+1) \cdot 2^{s-w}$ states~\cite{BuhlerKeichSunRECOMB03,kucherov2006ufs}.
On this automaton, our method runs in time $O(wm2^{s-w})$ which 
brings an improvement by a factor of
$\frac{2^w}{w}$ of the general bound $\mathcal{O}(m2^s)$ from \cite{BurkhardtKarkkainenFI03}.

\begin{center}%
\begin{figure}[!htb]%
\begin{minipage}[c]{0.5\textwidth} 
\begin{center}%
\includegraphics[width=0.95\textwidth]{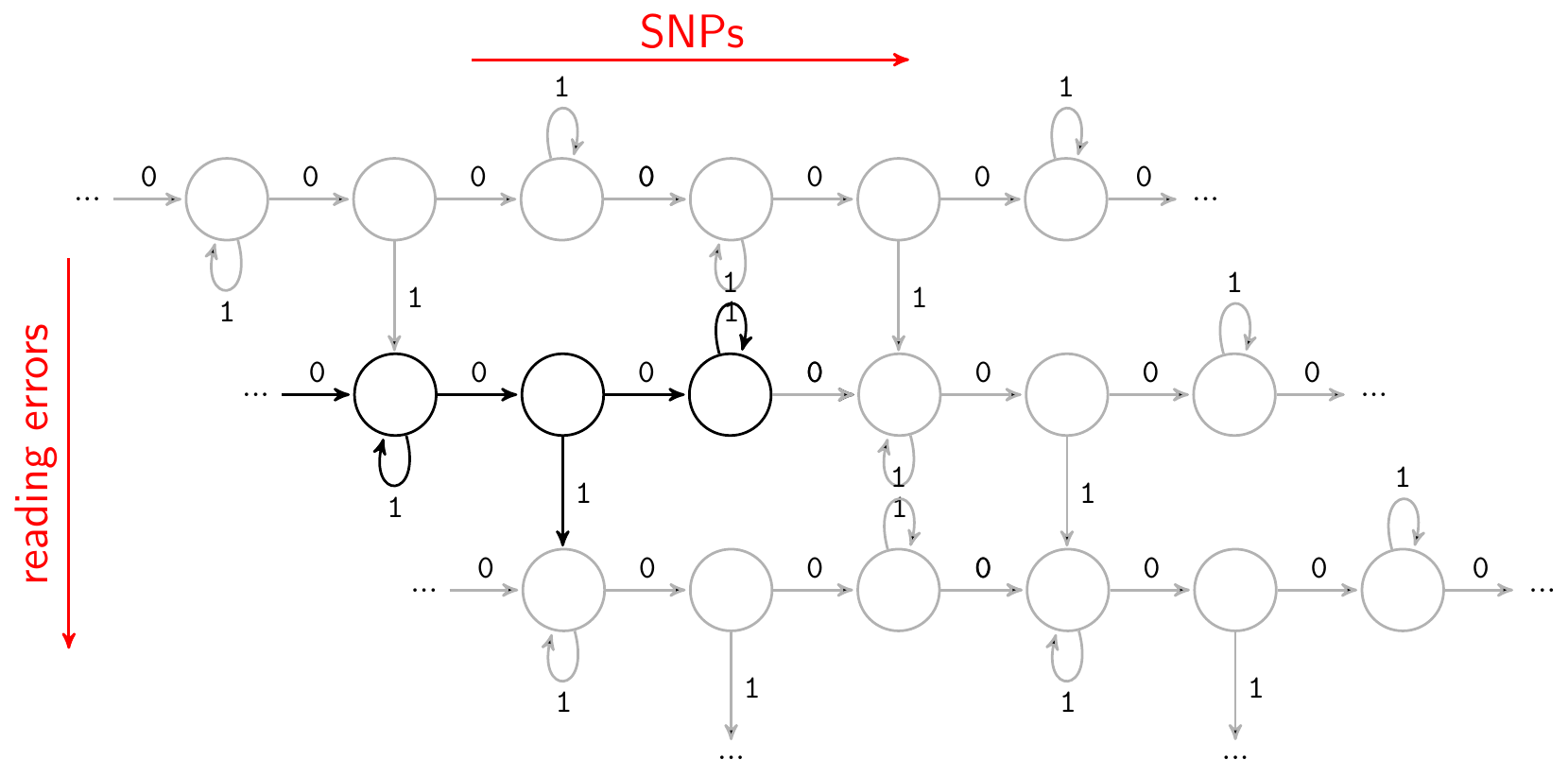}%
\vspace{-1em}
\caption{Building an automaton for  $k$ SNPs and $h$ color mismatches from
a repeated 3-state pattern.}%
\label{figure:1snp-1err-automaton-part}%
\end{center}%
\end{minipage}
\begin{minipage}[c]{0.5\textwidth} %
\begin{center}
\includegraphics[width=0.95\textwidth]{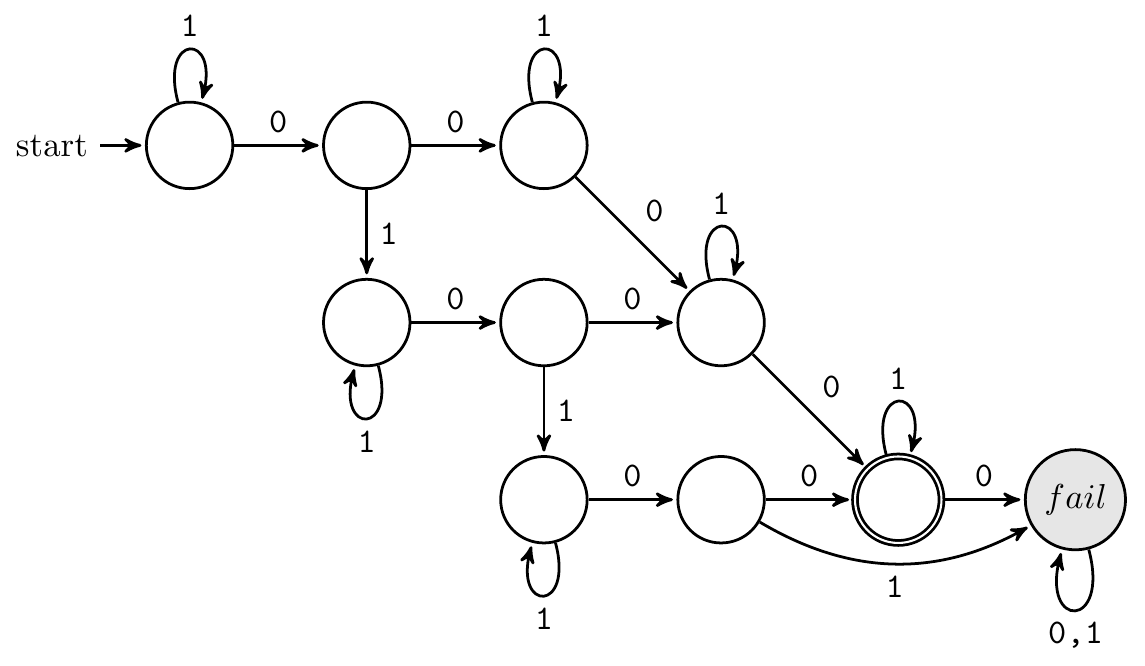}%
\vspace{-1em}
\caption{1 SNP \& 2 errors automaton.}%
\label{figure:1snp-2err-automaton}%
\end{center}
\end{minipage}
\end{figure}

\end{center}%

In the context of color sequence mapping, it is interesting
to define the lossless property with respect to a {\em maximal number of allowed
mismatches that is split between SNPs and reading errors}.
Since, in the color space, a SNP appears as two adjacent color mismatches,
having $k$ non-consecutive SNPs and $h$ color mismatches implies the possibility to
accept $2k + h$ mismatches with the additional restriction
that there exist at least $k$ pairs of adjacent ones.
The automaton that recognizes the set of alignments verifying
this condition on mismatches can be obtained by combining simple 3-state building
blocks as depicted in Figure~\ref{figure:1snp-1err-automaton-part}.
An example of such an automaton, accepting 1 SNP and 2 reading errors,
is illustrated in 
Figure~\ref{figure:1snp-2err-automaton} (1 and 0 denote match and
mismatch respectively).

Note that the case of consecutive SNPs, resulting in sequences of adjacent color mismatches,
is a simpler problem (since consecutive SNPs produce require less mismatches in the color
representation than the same number of non-consecutive SNPs)
and is covered by the proposed model:
a seed that is lossless for alignments with non-consecutive
SNPs will also be lossless for alignments with the same number of consecutive SNPs.

To verify the lossless property for $k$ SNPs and $h$ color mismatches,
we intersect the corresponding automaton with the seed automaton
(thus restricting the set of alignments recognized by the seed to those
with $k$ SNPs and $h$ color mismatches)
and submit the result to
the dynamic programming algorithm of Section~\ref{sec:eval-methods}.

\section{Experiments and Discussion}\label{sec:discussion}

We present now several efficient seed designs illustrating our
methodology (more examples at \url{http://bioinfo.lifl.fr/yass/iedera_solid}).

We first computed several sets of seeds of weight 10, restricted to
either 10 or 12 positions among the 34 positions of SOLiD reads,
each including one or two seeds. 
Figure~\ref{figure:seeds} shows some of the resulting seeds, together
with the corresponding sensitivity values, computed through the
methods described in Section~\ref{sec:seed-design}. 

\begin{figure}
\begin{center}
\begin{tabular}{ll}
\hline
~~~ {\small\sc 1-Lossy-10p:} sensitivity~0.9543 &~~~{\footnotesize\sc 2-Lossy-10p:} sensitivity~0.9627\\
\hline
~~~{\scriptsize\tt~1~~~5~~~~10~~~15~~~20~~~25~~~30~~~}~~~&~~~{\scriptsize\tt~1~~~5~~~~10~~~15~~~20~~~25~~~30~~~}~~~\\[-1mm]
~~~{\scriptsize\tt~\#\#\#\#\#\#\#--\#\#\#~~:~~~~:~~~~:~~~~:~~~~}~~~&~~~{\scriptsize\tt~\#\#\#\#\#\#\#\#\#\#~~~~:~~~~:~~~~:~~~~:~~~~}~~~\\[-2mm]
~~~{\scriptsize\tt~:~\#\#\#\#\#\#\#--\#\#\#:~~~~:~~~~:~~~~:~~~~}~~~&~~~{\scriptsize\tt~:~~~:~\#\#\#\#\#\#\#\#\#\#~~~:~~~~:~~~~:~~~~}~~~\\[-2mm]
~~~{\scriptsize\tt~:~~~\#\#\#\#\#\#\#--\#\#\#~~~:~~~~:~~~~:~~~~}~~~&~~~{\scriptsize\tt~:~~~:~~~~:~~\#\#\#\#\#\#\#\#\#\#~~:~~~~:~~~~}~~~\\[-2mm]
~~~{\scriptsize\tt~:~~~:~\#\#\#\#\#\#\#--\#\#\#~:~~~~:~~~~:~~~~}~~~&~~~{\scriptsize\tt~:~~~:~~~~:~~~~:~~~~\#\#\#\#\#\#\#\#\#\#:~~~~}~~~\\[-2mm]
~~~{\scriptsize\tt~:~~~:~~~\#\#\#\#\#\#\#--\#\#\#~~~~:~~~~:~~~~}~~~&~~~{\scriptsize\tt~:~~~:~~~~:~~~~:~~~~:~~~~\#\#\#\#\#\#\#\#\#\#}~~~\\[-2mm]
~~~{\scriptsize\tt~:~~~:~~~~:\#\#\#\#\#\#\#--\#\#\#~~:~~~~:~~~~}~~~&~~~{\scriptsize\tt~\#\#\#\#\#\#----\#\#\#\#:~~~~:~~~~:~~~~:~~~~}~~~\\[-2mm]
~~~{\scriptsize\tt~:~~~:~~~~:~~\#\#\#\#\#\#\#--\#\#\#:~~~~:~~~~}~~~&~~~{\scriptsize\tt~:~~~:~\#\#\#\#\#\#----\#\#\#\#~~~~:~~~~:~~~~}~~~\\[-2mm]
~~~{\scriptsize\tt~:~~~:~~~~:~~~~:\#\#\#\#\#\#\#--\#\#\#~~:~~~~}~~~&~~~{\scriptsize\tt~:~~~:~~~~:~~~\#\#\#\#\#\#----\#\#\#\#~~:~~~~}~~~\\[-2mm]
~~~{\scriptsize\tt~:~~~:~~~~:~~~~:~~~\#\#\#\#\#\#\#--\#\#\#~~~~}~~~&~~~{\scriptsize\tt~:~~~:~~~~:~~~~:~~\#\#\#\#\#\#----\#\#\#\#~~~}~~~\\[-2mm]
~~~{\scriptsize\tt~:~~~:~~~~:~~~~:~~~~:~~\#\#\#\#\#\#\#--\#\#\#}~~~&~~~{\scriptsize\tt~:~~~:~~~~:~~~~:~~~~:\#\#\#\#\#\#----\#\#\#\#}~~~\\
\hline
\end{tabular}
\begin{tabular}{ll}
\hline
~~~{\footnotesize\sc 1-Lossy-12p:} sensitivity~0.9626 &~~~{\footnotesize\sc 2-Lossy-12p:} sensitivity~0.9685 \\
\hline
~~~{\scriptsize\tt~1~~~5~~~~10~~~15~~~20~~~25~~~30~~~}~~~&~~~{\scriptsize\tt~1~~~5~~~~10~~~15~~~20~~~25~~~30~~~}~~~\\[-1mm]
~~~{\scriptsize\tt~\#\#\#\#\#\#\#--\#\#\#~~:~~~~:~~~~:~~~~:~~~~}~~~&~~~{\scriptsize\tt~\#\#\#\#\#\#\#\#\#\#~~~~:~~~~:~~~~:~~~~:~~~~}~~~\\[-2mm]
~~~{\scriptsize\tt~:~\#\#\#\#\#\#\#--\#\#\#:~~~~:~~~~:~~~~:~~~~}~~~&~~~{\scriptsize\tt~:~~~:~\#\#\#\#\#\#\#\#\#\#~~~:~~~~:~~~~:~~~~}~~~\\[-2mm]
~~~{\scriptsize\tt~:~~~\#\#\#\#\#\#\#--\#\#\#~~~:~~~~:~~~~:~~~~}~~~&~~~{\scriptsize\tt~:~~~:~~~~:\#\#\#\#\#\#\#\#\#\#~~~~:~~~~:~~~~}~~~\\[-2mm]
~~~{\scriptsize\tt~:~~~:~\#\#\#\#\#\#\#--\#\#\#~:~~~~:~~~~:~~~~}~~~&~~~{\scriptsize\tt~:~~~:~~~~:~~~~:\#\#\#\#\#\#\#\#\#\#~~~~:~~~~}~~~\\[-2mm]
~~~{\scriptsize\tt~:~~~:~~~\#\#\#\#\#\#\#--\#\#\#~~~~:~~~~:~~~~}~~~&~~~{\scriptsize\tt~:~~~:~~~~:~~~~:~~~~:\#\#\#\#\#\#\#\#\#\#~~~~}~~~\\[-2mm]
~~~{\scriptsize\tt~:~~~:~~~~:\#\#\#\#\#\#\#--\#\#\#~~:~~~~:~~~~}~~~&~~~{\scriptsize\tt~:~~~:~~~~:~~~~:~~~~:~~~~\#\#\#\#\#\#\#\#\#\#}~~~\\[-2mm]
~~~{\scriptsize\tt~:~~~:~~~~:~~\#\#\#\#\#\#\#--\#\#\#:~~~~:~~~~}~~~&~~~{\scriptsize\tt~\#\#\#\#--\#\#--\#\#\#\#:~~~~:~~~~:~~~~:~~~~}~~~\\[-2mm]
~~~{\scriptsize\tt~:~~~:~~~~:~~~~\#\#\#\#\#\#\#--\#\#\#~~~:~~~~}~~~&~~~{\scriptsize\tt~:~\#\#\#\#--\#\#--\#\#\#\#~~~:~~~~:~~~~:~~~~}~~~\\[-2mm]
~~~{\scriptsize\tt~:~~~:~~~~:~~~~:~\#\#\#\#\#\#\#--\#\#\#~:~~~~}~~~&~~~{\scriptsize\tt~:~~~\#\#\#\#--\#\#--\#\#\#\#~:~~~~:~~~~:~~~~}~~~\\[-2mm]
~~~{\scriptsize\tt~:~~~:~~~~:~~~~:~~~\#\#\#\#\#\#\#--\#\#\#~~~~}~~~&~~~{\scriptsize\tt~:~~~:~\#\#\#\#--\#\#--\#\#\#\#~~~~:~~~~:~~~~}~~~\\[-2mm]
~~~{\scriptsize\tt~:~~~:~~~~:~~~~:~~~~:\#\#\#\#\#\#\#--\#\#\#~~}~~~&~~~{\scriptsize\tt~:~~~:~~~\#\#\#\#--\#\#--\#\#\#\#~~:~~~~:~~~~}~~~\\[-2mm]
~~~{\scriptsize\tt~:~~~:~~~~:~~~~:~~~~:~~\#\#\#\#\#\#\#--\#\#\#}~~~&~~~{\scriptsize\tt~:~~~:~~~~:~~~~:~~~~:\#\#\#\#--\#\#--\#\#\#\#}~~~\\
\hline
\end{tabular}
\end{center}
\caption{Position-restricted seeds for 10 (above) and 12 (below)
  allowed positions. Different placements of a
seed correspond to the allowed positions.}\label{figure:seeds}
\end{figure}

Interestingly, both single seeds {\sc 1-Lossy-10p} and {\sc 1-Lossy-12p} contain a
double gap, which may reflect that an SNP modifies two adjacent colors. 
However, this gap is not
centered but rather shifted at the two-third of the seed (as observed
for the best single seeds of~\cite{kucherov2005mlf}).
Note also that in the two-seed families {\sc 2-Lossy-10p} and
{\sc 2-Lossy-12p}, one of the chosen
seeds is ungapped. This may be a consequence of the fact that we
consider indels in our lossy model, which usually forces the seeds to have a
smaller span.
Another interesting observation is that two-seed families {\sc
  2-Lossy-10p} and {\sc 2-Lossy-12p} 
are actually lossless for the threshold of 3 mismatches,
whereas single seeds {\sc 1-Lossy-10p} and
{\sc 1-Lossy-12p} are not lossless for this setting. 

We then focused on the lossless case where the maximal number of allowed
mismatches is split between SNPs and reading errors. Using the
procedure described in Section~\ref{sec:eval-methods},
we computed lossless single and double seeds for one SNP and two
reading errors.
Results are shown in Figure~\ref{figure:seeds-lossless-snp}.

\begin{figure}[!htb]
\begin{center}
\begin{tabular}{ll}
\hline
~~~~{\footnotesize \sc 1-Lossless-14p} & ~~~~{\footnotesize\sc 2-Lossless-8p}\\
\hline
~~~{\scriptsize\tt~1~~~5~~~~10~~~15~~~20~~~25~~~30~~~}~~~&~~~{\scriptsize\tt~1~~~5~~~~10~~~15~~~20~~~25~~~30~~~}~~~\\[-1mm]
~~~{\scriptsize\tt~\#\#\#\#--\#-------\#\#\#\#--\#~~~:~~~~:~~~~}~~~&~~~{\scriptsize\tt~\#\#\#\#\#\#\#\#\#\#~~~~:~~~~:~~~~:~~~~:~~~~}~~~\\[-2mm]
~~~{\scriptsize\tt~:\#\#\#\#--\#-------\#\#\#\#--\#~~:~~~~:~~~~}~~~&~~~{\scriptsize\tt~:~~~:~\#\#\#\#\#\#\#\#\#\#~~~:~~~~:~~~~:~~~~}~~~\\[-2mm]
~~~{\scriptsize\tt~:~\#\#\#\#--\#-------\#\#\#\#--\#~:~~~~:~~~~}~~~&~~~{\scriptsize\tt~:~~~:~~~~:~~~~:~~\#\#\#\#\#\#\#\#\#\#~~:~~~~}~~~\\[-2mm]
~~~{\scriptsize\tt~:~~\#\#\#\#--\#-------\#\#\#\#--\#:~~~~:~~~~}~~~&~~~{\scriptsize\tt~:~~~:~~~~:~~~~:~~~~:~~~\#\#\#\#\#\#\#\#\#\#~}~~~\\[-2mm]
~~~{\scriptsize\tt~:~~~~~...~~~~~~~~~~~~~~~:~~~~:}~~~&~~~{\scriptsize\tt~\#\#\#\#\#-----\#\#\#\#\#~~~~:~~~~:~~~~:~~~~}~~~\\[-2mm]
~~~{\scriptsize\tt~:~~~:~~~~:~~~\#\#\#\#--\#-------\#\#\#\#--\#}~~~&~~~{\scriptsize\tt~:\#\#\#\#\#-----\#\#\#\#\#~~~:~~~~:~~~~:~~~~}~~~\\[-2mm]
                                                                   &~~~{\scriptsize\tt~:~~~:~~~~:~~~~:~~\#\#\#\#\#-----\#\#\#\#\#~~}~~~\\[-2mm]
~~~{\scriptsize\tt(14 consecutive placements)}~~~&~~~{\scriptsize\tt~:~~~:~~~~:~~~~:~~~\#\#\#\#\#-----\#\#\#\#\#~}~~~\\
\hline
\end{tabular}
\caption{Lossless position-restricted seeds for 1 SNP and 2 reading errors}\label{figure:seeds-lossless-snp}

\end{center}
\end{figure}

Note that the seed {\sc 1-Lossless-14p} is one of several single seeds of
weight 10 we found that satisfied this lossless condition, with no
restriction on allowed positions. 
Interestingly, they all have a very large span (21) and a regular pattern
with a periodic structure that can be obtained by iterating a
simpler pattern solving the lossless problem for an appropriate {\em cyclic
  problem}, following the property we previously described in \cite{kucherov2005mlf}. 
For two-seed families, Figure~\ref{figure:seeds-lossless-snp} shows a
lossless pair of seeds {\sc 2-Lossless-8p} for read length 33 (which then
remains lossless for larger lengths), where each seed is
restricted to apply to four positions only. 

\begin{figure}
\begin{center}
\vspace{-2em}
\includegraphics[width=0.8\textwidth]{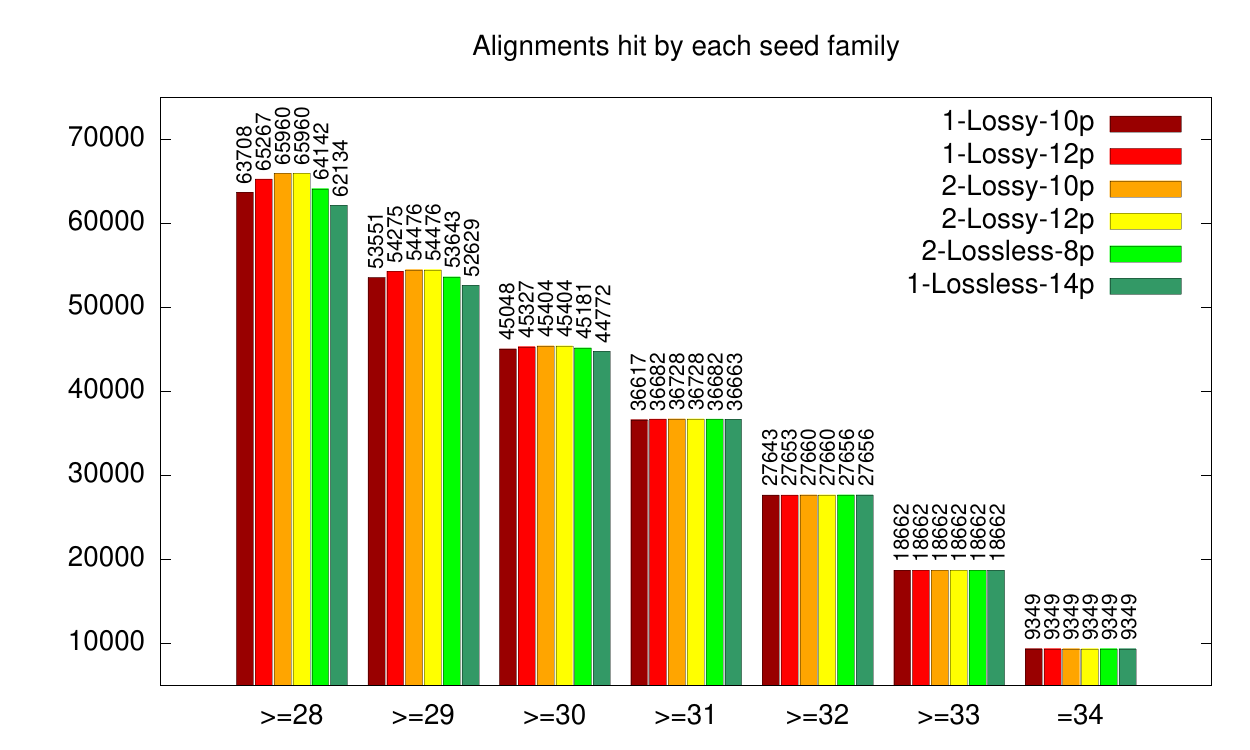}
\vspace{-2.5em}
\end{center}
\caption{Number of read alignments with scores between 28 and 34 hit by each seed}
\label{fig:score-bars}
\end{figure}

To get a better idea of the sensitivity of the obtained seeds
applied to real data, we tested them on 100000 reads
of length 34
from {\em S. cerevisiae} and computed the number
of read/reference alignments hit by each (single or double) seed. Alignments were
defined through the score varying from 28 to 34,
under the scoring scheme +1 for match,
0 for color mismatch or SNP, -2 for gaps. Results are presented
in Figure~\ref{fig:score-bars}.
One conclusion we can draw is that the performance of
lossless seeds 
{\sc 1-Lossless-14p} and  {\sc 2-Lossless-8p} 
decreases quite fast when the alignment score goes down, compared to lossy
seeds. Intuitively, this
is, in a sense, a price to pay for the lossless condition which
usually makes these seeds less appropriate for the alignments with a
number of errors exceeding the threshold. 
Another conclusion is that, as expected, single seeds perform worse than double seeds,
although the overall number of positions where seeds apply is the same
for both single and double seeds.

Note finally that the choice of the best seed
can be affected, on one hand, by different properties of the class of target alignments
(number, type and distribution of mismatches and
indels etc.) and, on the other hand, by the size of the data and the
available computational resources. The former can be captured by our
probabilistic models described in Section~\ref{sec:seed-design}. The
latter is related to the choice of the selectivity level, directly
affecting the speed of the search, which is defined by the seed weight
and the number of allowed positions. Depending on the chosen selectivity,
different seeds can (and should) be preferred. Note in this regard that seeds
appearing in Figure~\ref{fig:score-bars} have different selectivity
and are then incomparable {\em stricto sensu}. A comparison of
different seeds for SOLiD read mapping in typical practical
situations will be a subject of a separate work.

\section{Conclusions and perspectives}\label{sec:conclusions}

In this paper, we presented a seed design framework for mapping
SOLiD reads to a reference genomic sequence. 
Our contributions include the concept of position-restricted seeds, 
particularly suitable for short alignments with non-uniform error distribution;
a model that captures the statistical characteristics of the 
SOLiD reads, used for the evaluation of lossy seeds;
an efficient dynamic programming algorithm for verifying the
lossless property of seeds; the ability to distinguish between SNPs and
reading errors in seed design. 

Our further work will include a more rigorous training of our models
and in particular a more accurate estimation of involved
probabilities, possibly using advanced methods of assessing the fit of
a model. Another interesting question to study is the design of
efficient combined lossy/lossless seeds which provide a guarantee to
hit all the alignments with a specified number of errors and still
have a good sensitivity when this threshold is exceeded. Computing
such seeds, however, could be difficult or even unfeasible: for
example, lossless seeds tend to have a regular structure (see
\cite{kucherov2005mlf}) while best lossy seeds often have asymmetric
and irregular structure. Finally, we want to define and study a lossless
property that incorporates possible indels and not only mismatches (SNPs
or reading errors) occurring in read alignments.

\section*{Acknowledgments}\label{sec:ack}

The authors would like to thank Valentina Boeva and Emmanuel Barillot
from the {\em Institut Marie Curie} at Paris for helpful discussions
and for providing the dataset of {\em Saccharomyces cerevisiae} reads
that we used as a testset in our study. 
We also thank Martin Figeac ({\em Institut national de la
  sant\'e et de la recherche m\'edicale}) for sharing 
insightful knowledge about the SOLiD technology.
Laurent No{\'e} was supported by the ANR project CoCoGen (BLAN07-1 185484).

\bibliographystyle{splncs}
\bibliography{maps.bib}

\end{document}